\documentclass{article}
\pdfoutput=1
\usepackage{trcover}

\title{Toward a Formal Semantics for Autonomic Components}
\trnum{TR-08-08}
\author{Marco Aldinucci \and Emilio Tuosto\thanks{Department of Computer Science, University of Leicester, UK}}
\date{April 22, 2008}

\usepackage{xspace,graphicx,url}
\usepackage{booktabs,layouts}

    \usepackage{picinpar}
    \usepackage{amsmath}
    \usepackage{amssymb}
    \usepackage{graphicx}
    \usepackage[all]{xy}
    \CompileMatrices
    \usepackage{boxedminipage}
    \usepackage[usenames]{color}
    \usepackage{multicol}
    \usepackage{pstricks,pst-node,pst-text,pst-3d}
    \usepackage{subfigure}
    \usepackage{epsfig}
    \usepackage{ifthen}
    \usepackage{rotating}
    \usepackage{fancybox}
    \usepackage{calc}
    \usepackage{paralist}

 \newif\ifemi
    \emitrue
    \emifalse
  
  \newif\ifthm
    \thmfalse
    \thmtrue
  
  \newif\ifllncs
    \llncstrue
    \llncsfalse
  
  \newif\ifarticle
    \articlefalse
    \articletrue
  
  \newif\iftesi
    \tesitrue
    \tesifalse

  \newcommand{\comment}[1]{}

  \newcommand{\proofend}{\mbox{$\Box$}}

  \newcommand{\nar}{\ar@{-}}
  
  \newcommand{\envprod}[3]{\xymatrix@R1.7pc{
    \bullet
      \POS[]+<0pc,-.5pc>\drop{x}
      \POS[]+<.0pc,.5pc>\drop{\mbox{\small$#1$}}
    & *+[F]{\env}\nar[l]\nar[d]\nar[r]
    & \bullet
        \POS[]+<0pc,-.5pc>\drop{y}
        \POS[]+<0pc,.5pc>\drop{\mbox{\small$#2$}}
    & \Longrightarrow
    & \bullet
        \POS[]+<0pc,-.5pc>\drop{x}
    & *+[F]{\env}\nar[l]\nar[d]\nar[r] &
      \bullet
      \POS[]+<0pc,-.5pc>\drop{y}
    \\
    & \bullet
        \POS[]+<-.5pc,0pc>\drop{z}
        \POS[]+<2.5pc,0pc>\drop{\mbox{\small$#3$}}
    &
    & &
    & \bullet\POS[]+<-.5pc,0pc>\drop{z}
  }}

  \newcommand{\strans}[1]{\xymatrix{\ar@{~>}[r]^{#1} &}}
  \newcommand{\dstrans}[1]{\xymatrix{\ar@{~>}[d]^{#1} \\ \ }}
  
  \newcommand{\arw}[1]{\overstackrel{\rightarrowfill}{#1}}

\newcommand{\conf}[1]{\langle {#1} \rangle}

 \newtheorem{example}{Example}[section]

\newcommand{\env}{E_d}

\def \overunderstackrel#1#2#3{\mathrel{\mathop{#1}\limits^{#2}_{#3}}}
\def \overstackrel#1#2{\mathrel{\mathop{#1}\limits^{#2}}}

\def\act#1#2{#1 @ #2}

\newcommand{\trans}[3]{\mbox{$#1 \stackrel{{#2}}{\rightarrowfill} {#3}$}}

  \input{prooftree}

 \ifemi
    \newcommand{\emi}[1]{{\marginpar[\scriptsize {#1}]{\scriptsize {#1}}\label{:emi:}}}       
  \else
    \newcommand{\emi}[1]{}
  \fi
 
\begin{document}
 \maketitle





  \begin{abstract}
Autonomic management can improve the QoS provided by parallel/ distributed
applications. Within the CoreGRID Component Model, the autonomic
management is tailored to the automatic -- monitoring-driven  -- 
alteration of the component assembly and, therefore, is defined as the effect
of (distributed) management code.

This work yields a semantics based on \emph{hypergraph} rewriting suitable
to model the dynamic evolution and non-functional aspects of Service
Oriented Architectures and component-based autonomic applications. In
this regard, our main goal is to provide a formal description of adaptation
operations that are typically only informally specified.
We contend that our approach makes easier to raise the level of
abstraction of management code in autonomic and adaptive applications.
\end{abstract}
  
  \section{Introduction}
\label{into:sec}

Developers of grid applications cannot rely neither on fixed
target platforms nor on stability of their status \cite{grads:overview}.
This makes dynamic adaptivity of applications an essential feature in
order to achieve user-defined levels of Quality of Service (QoS).
In this regard, component technology has gained increased impetus in
the grid community for its ability to provide a clear separation of
concerns between application logic and QoS-driven adaptation, which
can also be achieved \emph{autonomically}.
As an example, GCM (the Grid Component Model defined within the
CoreGRID NoE) is a hierarchical component model explicitly designed to
support component-based autonomic applications in highly dynamic and
heterogeneous distributed platforms \cite{gcm:coregrid:07}.


An assembly of components may be naturally modeled as a graph and, if
components are autonomic, the graph can vary along with the program
execution and may change according to input data and/or grid hardware
status.
These changes can be encoded as reaction rules within the component
\emph{Autonomic Manager} (hereafter denoted as $AM$).
A proper encoding of these rules effectively realises the management
policy, which can be specific of a given assembly or pre-defined for
parametric assemblies (such as \emph{behavioural skeletons})
\cite{orc:pdp:08,advske:pc:06}.
In any case, the management plan relies on the reconfiguration
operation exposed by the component model run-time support.

A major weakness of current component models
(including GCM) is that the semantics of these operations are
informally specified, thus making hard to reason about QoS-related
management of components.  In this work
\smallskip
\begin{compactitem}
\item We discuss few primitives useful for component adaptation; the
chosen operations are able to capture typical adaptation patterns in
parallel/ distributed application on top of the grid.  These are
presented as \emph{non-functional interfaces} of components that
trigger component assembly adaptation.
\item We detail a semantics for these operations based on
\emph{hypergraph} rewriting suitable for the description of component
concurrent semantics and assembly evolution along adaptations.
\end{compactitem}
\smallskip
The key idea of our semantical model consists in modeling component-based
applications by means of \emph{hypergraphs} which generalise usual graphs
be allowing \emph{hyperedges}, namely arcs that can connect more than two
nodes.
Intuitively, hyperedges represent components able to interact through
\emph{ports} represented by nodes of hypergraphs.
The \emph{Synchronized Hyperedge Replacement} (SHR) model specifies
how hypergraphs are rewritten according to a set of \emph{productions}.
Basically, rewritings represent adaptation of applications possibly
triggered by the underlying grid middleware events (or by the applications
themselves).

SHR has been shown suitable for modelling non-functional aspects of
service oriented computing~\cite{dfmpt03,fhlmt06}\ and is one of the modelling
and theoretical tools of the {\sc Sensoria} project~\cite{Sensoria}.
For simplicity, we consider a simplified version of SHR where node fusions is
limited and restriction is not considered.
Even if, for the sake of simpleness, the SHR framework used in this
work is not the most general available, it is sufficient to give semantics
to the management primitives (aka adaptation operations) addressed
here. The autonomic manager --  by way of these adaptation operations -- can
structurally reconfigure an application to pursue the (statically or
dynamically specified) user intentions in terms of  QoS.


%


\section{Autonomic Components and GCM}
\label{sec:gcm}
Autonomic systems enables dynamically defined adaptation by allowing
adaptations, in the form of code, scripts or rules, to be added,
removed or modified at run-time. These systems typically rely on a
clear separation of concerns between adaptation and application logic
\cite{AC:vision:2003}. An autonomic component will typically consist
of one or more managed components coupled with a single autonomic
manager that controls them.  To pursue its goal, the manager may
trigger an adaptation of the managed components to react to a run-time
change of application QoS requirements or to the platform status. In
this regard, an assembly of self-managed components implements, via
their managers, a distributed algorithm that manages the entire
application.

The idea of autonomic management of parallel/distributed/grid
applications is present in several programming frameworks, although in
different flavours: ASSIST \cite{van:assist:02,advske:pc:06}, AutoMate
\cite{ac:automate:06}, SAFRAN \cite{ac:safran:06}, and GCM
\cite{gcm:coregrid:07}\ all include autonomic management features. The
latter two are derived from a common ancestor, i.e. the Fractal
hierarchical component model \cite{fractal:spec}. All the named
frameworks, except SAFRAN, are targeted to distributed applications on
grids.

GCM builds on the Fractal component model
\cite{fractal:spec}\ and exhibits three prominent features: hierarchical composition, collective interactions and autonomic management.
GCM components have two kinds of interfaces:
functional and non-functional ones. The functional interfaces host
all those ports concerned with implementation of the functional features of
the component.
The
non-functional interfaces host all those ports needed to support the
component management activity in the implementation of the
non-functional 
features, i.e. all those features contributing to the efficiency of
the component in obtaining the expected (functional) results
but not directly involved in result computation. Each GCM
component therefore contains an $AM$, 
interacting with other managers in other components via the component
non-functional interfaces. The $AM$ implements the autonomic cycle via
a simple program based on reactive rules. 
%
These rules are typically specified as a collection of
\texttt{when-}\emph{event}\texttt{-if-}
\emph{cond}\texttt{-then-}\emph{adapt\_op}  
clauses, where \emph{event} is raised by the monitoring of
component internal or external activity (e.g. the component server
interface received a request, and the platform running a component
exceeded a threshold load, respectively); \emph{cond} is an expression
over component internal attributes (e.g. component life-cycle status);
\emph{adapt\_op} represents an adaptation operation (e.g. create, destroy a
component, wire, unwire components, notify events to another
component's manager) \cite{ac:safran:06}.

We informally describe some common adaptation operations that
may be assigned to configuration interfaces are the following:
\smallskip
\begin{compactdesc}
\item[Migration] A component is required to change its running
  location (e.g. platform, site). The request must include the new
  location and can be 
  performed while keeping its attached external state ($\mathbf{go}$) or
  restating from a fresh default state ($\mathbf{start}$).
\item[Replication]
  A component (either composite or primitive) is
  replicated. Replication operation is particularly targeted to
  composite components exhibiting the parametric replication of inner
  components (such as behaviour skeletons), and can be used to change their
  parallelism degree (and thus their performance and fault-tolerance
  properties).   
  Replication events are further characterized with respect to their
  relation with replicated component state, if any. 
  A component replica may be created with a fresh external state,
  carry a copy of the external state 
  ($\mathbf{copy}$), or share the external state with the source component
  ($\mathbf{share}$).
\item[Kill] A component is killed. Due to this kind of action
  disconnected components (and in particular storage managers) can
  subject to garbage collection.
\end{compactdesc}
\smallskip
Described primitives make possible the implementation of several
adaptation paradigms. In particular, migration may be used to
adapt the application to changes of grid topology as well as to
performance drop of resources. Replication and kill may be used to adapt
both data and task parallel computation. In particular, replication
with share enables the redistribution of sub-task in data
parallel computations; replication with copy enables
hot-redundancy. Both stateful and stateless farm computation
(parameter-sweeping, embarrassingly parallel) 
may be reshaped both in parallelism degree and location run by using
replication and kill.  

\begin{example}
\label{es:global}
Let \emph{P, C, SF, S, $AM$, W${}_1$, W${}_2$, W${}_3$} components
(Producer, Consumer, Stateful Farm\footnote{This
  component is a  composite component, and in particular it is an
  instance of a behavioural skeleton \cite{orc:pdp:08}.}, Storage, Autonomic Manager, and
Workers); \emph{L}${}_1 \cdots$ \emph{L}${}_8$ locations. Thee kinds of bindings are used  in the assembly (see also
Sec.~\ref{shrgrid:sec}). \\[1ex]
 \centerline{\includegraphics[width=\linewidth]{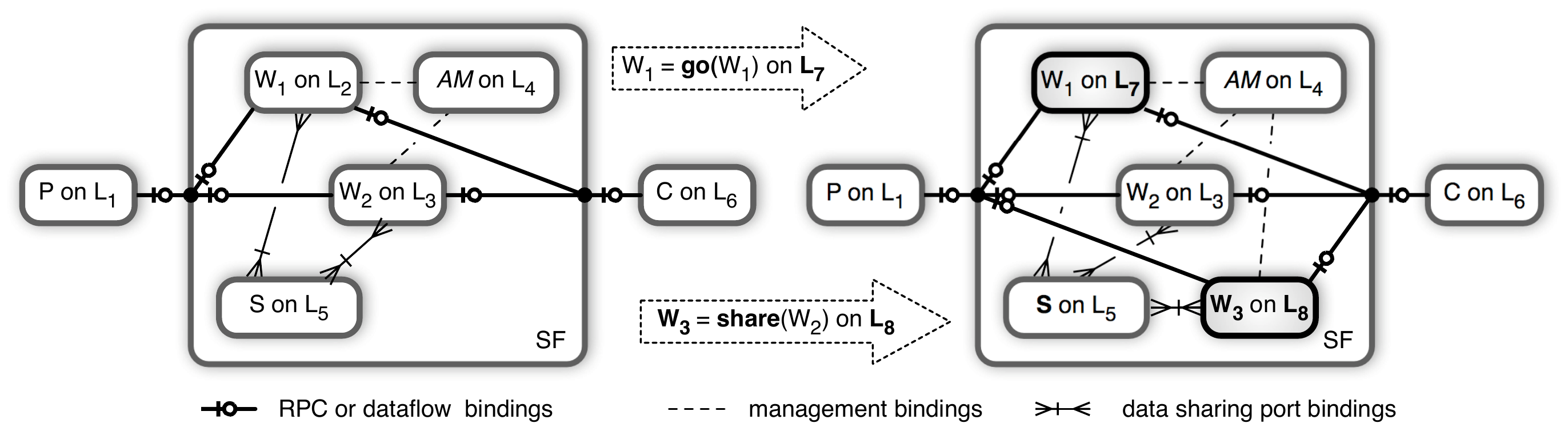}}\\[1ex]
The described assembly of components (left) is paradigmatic of many
producer-filter-consumer applications, where the producer (P)
generates a stream of data and the filter is parallel component (SF)
exhibiting a shared state among its inner components (e.g. a
database). The original assembly (left) can be dynamically adapted
(right) by way of two adaptation operations to react to run-time
events, such as a request of increasing the throughput. The
\emph{\textbf{go}} operation moves \emph{W}${}_1$ from \emph{L}${}_2$ to
\emph{L}${}_7$ (as an example to move a component onto a more powerful
machine); the \emph{\textbf{share}} operation that replicates
\emph{W}${}_2$ and place it in the new location \emph{L}${}_8$ (to
increase the parallelism degree). Both
operations preserve the external state of the migrated/replicated
component, which is realised by way of a storage component) attached
via a data sharing interface \cite{AntBouJanPer06}. 
\end{example}

Example \ref{es:global} illustrates how the management can be described from a
\emph{global viewpoint}.  Indeed, the system is described by in a
rather detailed way, e.g., components are explicitly enumerated along
with their connections.  Even if this global viewpoint is useful (and
sometime unavoidable) when designing distributed systems, it falls
short in describing what single components are supposed to do when a
reconfiguration is required.  In other terms, it is hard to tell what
the \emph{local} behaviour of each component should be in order to
obtain the reconfiguration described by the \emph{global} view.

Also, it is worth remarking that, though the diagram clearly describes
the changes triggered by $AM$ in this scenario, the lack of a formal
semantics leaves some ambiguities.  For example, it is not clear if
the reconfiguration should take place if, and only if, the system is
configured as on the lhs or this is rather a ''template''
configuration (e.g., should the system reconfigure itself also when
$W_2$ is connected to $W_1$ rather than to $D$? What if $W_2$ was not
present?).  Of course, such ambiguous situations can be avoided when a
formal semantics is adopted.



  \newcommand{\confint}{non-functional interface}
\newcommand{\Confint}{Non-functional interface}
\newcommand{\ConfInt}{Non-functional Interface}
\newcommand{\gridnode}{\diamond}
\newcommand{\storenode}{\bullet}
\newcommand{\store}{\sigma}
\newcommand{\ggo}{\mathbf{go}}
\newcommand{\gstart}{\mathbf{start}}
\newcommand{\gbang}{\mathbf{rep}}
\newcommand{\gcopy}{\mathbf{copy}}
\newcommand{\gkill}{\mathbf{kill}}

\section{A Walk through SHR}
\label{hgintro:sec}
\emph{Synchronised Hyperedge Replacement} (SHR) can be thought of as a
rule-based framework for modelling (various aspects of) of distributed
computing~\cite{fhlmt06}\ modelled as \emph{hypergraphs}, a generalisation of
graphs roughly representing (sets of) relations among nodes.
While graphs represent (sets of) binary relations
(labelled arcs connect exactly two nodes), labelled \emph{hyperedges}
(hereafter, edges) can connect any number of nodes.
We give an informal albeit precise description of \emph{hypergraphs}
and SHR through a suitable graphical notation.
The interested reader is referred to~\cite{fhlmt06,lt05}\ and
references therein for the technical details.
\begin{example}\label{ex:hg}
  In our graphical notation, a hypergraph is depicted as
  \[\xymatrix@R.7pc@C1.5pc{
    \bullet\POS[]+<.7pc,0pc>\drop{l} & &
    *+[F]{AM}
    \nar@(r,ul)[rr]
    \nar@(l,r)[lld]
    & & \bullet\POS[]+<.3pc,.7pc>\drop{l'}
    & \bullet\POS[]+<0pc,.7pc>\drop{s} & *+[F]{\store}\nar[l]
    \\
    \bullet\POS[]+<0pc,-.7pc>\drop{g}
    & & *+[F]{f}
    \nar[ll]
    \nar@(r,l)[urr]
    \nar@(r,dl)[rr]
    & & \bullet\POS[]+<.3pc,-.7pc>\drop{s'} & *+[F]{\store}\nar[l] 
  }\]
  Edges (labelled by $f$, $AM$ and $\store$) are connected to
  nodes ($g$, $l$, $l'$, $s$ and $s'$).
  Specifically, $AM$ connects $g$ and $l'$,
  $f$ connects $g$, $l'$ and $s'$ while two $\store$-labelled edges
  are attached to $s$ and $s'$.
  Notice that nodes can be isolated (e.g., $l$).
\end{example}
Hyperedges represent (distributed) components that interact through \emph{ports}
represented by nodes.
Connections between edges and nodes, called \emph{tentacles}, allow
components sharing ports to interact (e.g., in Example~\ref{ex:hg}, $f$
and $AM$ can interact on $g$ and on $l'$).
\begin{example}
  The hypergraph in Example~\ref{ex:hg} represents (part of) a system where
  a manager $AM$ and a component $f$ are located at $l'$ and can interact on
  port $g$.
  The component $f$ has access to the store at $s'$ (e.g. by way of a
  data port \cite{AntBouJanPer06}). 
  In the system are also present another location $l$ and store $s$.
\end{example}

As in string grammars, SHR rewriting is driven by \emph{productions}.
In fact, strings can be rewritten according to a set of
\emph{productions}, i.e. rules of the form $\alpha \arw{} \beta$,
where $\alpha$ and $\beta$ are strings (over fixed alphabets of
terminal and non-terminal symbols).
Similarly, in SHR hypergraph rewritings are specified by productions of
the form $\trans{L}{}{R}$, where the lhs $L$ is a hyperedge, the rhs
$R$ is a hypergraphs and states that occurrences of $L$
can be replaced with $R$.
Intuitively, edges correspond to non-terminals and can
be replaced with a hypergraph according to their productions.
In SHR, hypergraphs are rewritten by \emph{synchronising} productions,
namely edge replacement is \emph{synchronised}: to apply the
productions of edges sharing nodes, some conditions must be fulfilled.
More precisely, an SHR production can be represented as follows:

{\small\label{page:prod}\[\begin{array}{c|c}
  \xymatrix@R1pc{
    & & \bullet\POS[]+<-.7pc,0pc>\drop{l}
    \\
    \bullet\POS[]+<-.7pc,0pc>\drop{g}
    & & *+[F]{f}
    \nar_{\gcopy \conf{g',s',l'}}[ll]
    \nar[u]
    \nar_{\overline{\gbang} \conf{s'}}[rr]
    & & \bullet\POS[]+<.7pc,0pc>\drop{s}
  }
  \hspace{1cm}&\hspace{1cm}
  \xymatrix@R1pc@C1pc{
    \bullet\POS[]+<-.7pc,0pc>\drop{g'}
    & *+[F]{f}
    \nar[l]
    \nar@/^.5pc/[rr]
    \nar@/^1pc/[rrr]
    & \bullet\POS[]+<-.7pc,0pc>\drop{l}
    & \bullet\POS[]+<.7pc,0pc>\drop{l'}
    &  \bullet\POS[]+<0pc,.7pc>\drop{s'} &
    \\
    \bullet\POS[]+<-.7pc,0pc>\drop{g}
    & & 
*+[F]{f}
    \nar[ll]
    \nar[u]
    \nar[rr]
    & & \bullet\POS[]+<.7pc,0pc>\drop{s}
  }
\end{array}\]}
where on the lhs is a decorated edge and on the rhs a hypergraph.
The production above should be read as a rewriting rule specifying that edge
$f$ on the lhs can be replaced with the hypergraph on the rhs provided that
the conditions on the tentacles are fulfilled.
More precisely, $\gcopy$ and $\overline\gbang$ must be satisfied on node $g$ and
$s$, respectively while $f$ is \emph{idle} on node $l$, namely it does not pose
any condition on $l$.
According to our interpretation, this amounts to say that when component $f$ is
said to replicate with copy by its $AM$ (condition $\gcopy$ on node $g$), it tells
its store to duplicate itself (condition $\gbang$ on node $s$).
When such conditions are fulfilled, edge $f$ is replaced with the hypergraph on the
rhs which yield two instances of $f$ one of which connected to the communicated
nodes as prescribed by the rhs of the production.
Indeed, $f$ exposes three nodes on condition $\gcopy$ and one on
$\gbang$; these represent nodes that are communicated (i.e., $g$ and $l$ are
node communication accounts for mobility as edges can dynamically detach
their tentacles from nodes and connect them elsewhere.

SHR has a declarative flavour because programmers specify synchronization
conditions of components independently from each other.
Once the system is built (by opportunely connecting its components) it will
evolve according to the possible synchronizations of the edges.
Global transitions are obtained by parallel application of productions
with ``compatible'' conditions where compatibility depends on the chosen
synchronisation policy\footnote{
  SHR is parametric with respect to the
  synchronisation mechanism adopted and can even encompass several
  synchronisation mechanisms~\cite{fhlmt06,lt05}.
}.

\comment{
Conditions on $\trans{L}{}{R}$ allow one to introduce the concept of
``context-freeness'': More precisely, productions
with a left-hand-side (lhs) which is either a node or an edge confer a
``context-free'' flavor to graph grammars.
Indeed, such productions do not consider the ``surroundings'' of their
lhs.

Hypergraphs are graphs made of nodes and a generalization of edges called
\emph{hyperedges}, in the sense that hyperedges may connect more/less
than two nodes.

  An edge in a traditional graph can be intuitively thought of as representing
  a binary relation between two nodes, while a hyperedge represents a relationships
  among many nodes.

Figure~\ref{hg-replacement:fig}(a) depicts an hyperedge $f$ relating five nodes.

We will describe a rewriting mechanism based on hyperedge replacement via synchronization;
indeed, such mechanism will be exploited to constraining graph rewriting.
Figure~\ref{hg-replacement:fig} aims at giving an intuition of hyperedge replacement.
\begin{figure}\center
  \begin{tabular}{cc}
    \subfigure[An hyperedge $f$]{
      \includegraphics[width=5cm]{./images/graph-rw1.eps}
    }
    & \qquad
    \subfigure[Replacement of $f$ with $G$]{
      \includegraphics[width=5cm]{./images/graph-rw2.eps}
    }
  \end{tabular}
  \caption{Hyperedge replacement}
  \label{hg-replacement:fig}
\end{figure}
The edge $f$ in Figure~\ref{hg-replacement:fig}(a) is connected to $G1$ and $G2$ through
its \emph{external nodes} (or \emph{attachment points}) $\mathbf{1},\ldots,\mathbf{5}$:
Figure~\ref{hg-replacement:fig}(b) represents the hypergraph obtained by replacing $f$
with hypergraph $G$.
The dashed gray part of the Figure~\ref{hg-replacement:fig}(b) represents the initial
situation and disappears after $f$ has been replaced by $G$; notice that $G_1$ is not involved
in the rewriting.
Moreover, new nodes can appear (all nodes in $G$
but $\mathbf{1}$ and $\mathbf{2}$ are new nodes generated by the transition) and some
nodes can be ``fused'' after the transition (in Figure~\ref{hg-replacement:fig}(b) node
$\mathbf{4}$ is fused with $\mathbf{5}$).
As will be shown later, this amounts to {\em mobility\/} of components, that
dynamically can change their connections.
In fact, notice that, in Figure~\ref{hg-replacement:fig}(b), the part of $G2$
that was connected to node $\mathbf{4}$ corresponds, in Figure~\ref{hg-replacement:fig}(a),
to the part of $G2'$ connected to node $\mathbf{5}$.
}
  \section{Productions for \ConfInt s}
\label{shrgrid:sec}
SHR can adequately formalise the \confint\
mechanisms informally described in Sec.~\ref{sec:gcm}.
Three conceptually distinct interfaces can be considered:
$i)$ interfaces between components and $AM$ (for management bindings),
$ii)$ interfaces toward the external state (for data sharing bindings), and
$iii)$ interfaces for communicating with other components (for
RPC/dataflow bindings).
Since interfaces $iii$ are application dependent, we
focus on the coordination-related interfaces $i$ and $ii$.

A main advantage of our approach is that all aspects of \confint s are captured
in a uniform framework based on SHR.
Indeed,
\smallskip
\begin{compactitem}
  \item components are abstracted as edges connected to
     form a hypergraph;
  \item the coordination interface of each component is separately declared and
     is not mingled with its computational activity;
  \item being SHR a \emph{local} rewriting mechanism, it is possible to specify
     confined re-configuration of systems triggered by \emph{local} conditions;
\end{compactitem}

\paragraph{Migration}
The migration of a component $f$ is triggered when its $AM$ raises a signal
$\ggo$ with the new location on node $g$.
The synchronisation of $f$ on the $\ggo$ signal is given by following
production:

{\small\[\begin{array}{c|c}
  \xymatrix@R1pc@C1.5pc{
    & & \bullet\POS[]+<-.7pc,0pc>\drop{l}
    \\
    \bullet\POS[]+<-.7pc,0pc>\drop{g}
    & & *+[F]{f}
    \nar_{\ggo\conf{g',l'}}[ll]
    \nar[u]
    \nar[rr]
    & & \bullet\POS[]+<-0pc,.7pc>\drop{s}
  }
  \hspace{1cm}&\hspace{1cm}
  \xymatrix@R1pc@C.7pc{
    \bullet\POS[]+<-.7pc,0pc>\drop{g'} & & \bullet\POS[]+<-.7pc,0pc>\drop{l} & \bullet\POS[]+<.7pc,0pc>\drop{l'}
    \\
    \bullet\POS[]+<-.7pc,0pc>\drop{g}
    & & *+[F]{f}
    \nar@(l,dr)[ull]
    \nar@(ur,dl)[ur]
    \nar[rr]
    & & \bullet\POS[]+<.7pc,0pc>\drop{s}
  }
\end{array}\]}
specifying that $f$ running at $l$ accepts to migrate to $l'$ (lhs);
the ``location'' tentacle of $f$ is disconnected from $l$ and attached
to $l'$ (rhs).
Notice that $f$ maintains the connection to the previous state $s$ and
$l$ is still present.
The tentacle connected to $g$ on the lhs is connected to $g'$ on the rhs;
however, it might well be that $g'=g$ ($f$ is still connected to the
original $AM$) or $g \neq g'$ ($f$ changes manager).
Similarly, $\gstart$ moves the component to a new location $l'$.
However, a new external state $\store$ is created together with its
attaching node:

{\small\[\begin{array}{c|c}
  \xymatrix@R1pc@C1.5pc{
    & & \bullet\POS[]+<-.7pc,0pc>\drop{l}
    \\
    \bullet\POS[]+<-.7pc,0pc>\drop{g}
    & & & *+[F]{f}
    \nar|-{\gstart\store\conf{g',l',s'}}[lll]
    \nar[ul]
    \nar[r]
    & \bullet\POS[]+<-0pc,.7pc>\drop{s}
  }
  \hspace{1cm}&\hspace{1cm}
  \xymatrix@R1pc@C1pc{
    \bullet\POS[]+<-.7pc,0pc>\drop{g'} & & \bullet\POS[]+<-.7pc,0pc>\drop{l}
    & \bullet\POS[]+<.7pc,0pc>\drop{l'}
    & & *+[F]{\store}\nar[d]
    \\
    \bullet\POS[]+<-.7pc,0pc>\drop{g}
    & & *+[F]{f}
    \nar[ull]
    \nar[ur]
    \nar@/_1pc/[rrr]
    & & \bullet\POS[]+<-0pc,.7pc>\drop{s} & \bullet\POS[]+<-.7pc,.7pc>\drop{s'}
  }
\end{array}\]}

\paragraph{Replication}
Unlike migration, replication of $f$ preserves its location:

{\small\[\begin{array}{c|c}
  \xymatrix@R1pc@C1.5pc{
    & & \bullet\POS[]+<-.7pc,0pc>\drop{l}
    \\
    \bullet\POS[]+<-.7pc,0pc>\drop{g}
    & & & *+[F]{f}
    \nar|-{\gbang\conf{g',l'}}[lll]
    \nar[ul]
    \nar[r]
    & \bullet\POS[]+<.7pc,0pc>\drop{s}
  }
  \hspace{1cm}&\hspace{1cm}
  \xymatrix@R1pc@C1pc{
    \bullet\POS[]+<-.7pc,0pc>\drop{g'} & & \bullet\POS[]+<-.7pc,0pc>\drop{l}
    & \bullet\POS[]+<-.7pc,0pc>\drop{l'}
    & & *+[F]{f}
    \nar@/_1pc/[lllll]
    \nar[d]
    \nar[ll]
    \\
    \bullet\POS[]+<-.7pc,0pc>\drop{g}
    & & & *+[F]{f}
    \nar[lll]
    \nar[ul]
    \nar[rr]
    & & \bullet\POS[]+<.7pc,0pc>\drop{s}
  }
\end{array}\]}
the effect of the above production is to add a new instance of $f$ at $l'$
with $AM$ connected to $g'$; of course, $l=l'$ and $g=g'$ are possible.
The newly generated instance shares external state with the original one.

Replication can also activate the new instance with a different state:

{\small\[\begin{array}{c|c}
  \xymatrix@R1pc@C1.5pc{
    & & \bullet\POS[]+<-.7pc,0pc>\drop{l}
    \\
    \bullet\POS[]+<-.7pc,0pc>\drop{g}
    & & & *+[F]{f}
    \nar|-{\gbang\store\conf{g',l'}}[lll]
    \nar[ul]
    \nar[r]
    & \bullet\POS[]+<.7pc,0pc>\drop{s}
  }
  \hspace{1cm}&\hspace{1cm}
  \xymatrix@R1pc@C1pc{
    *+[F]{f}
    \nar[d]
    \nar@/^.5pc/[rrr]
    \nar@(u,ur)[rrrrrd]
    & & \bullet\POS[]+<-.7pc,0pc>\drop{l}
    & \bullet\POS[]+<.7pc,0pc>\drop{l'}
    \\
    \bullet\POS[]+<-.7pc,0pc>\drop{g}
    & & & 
*+[F]{f}
    \nar[lll]
    \nar[ul]
    \nar[rr]
    & & \bullet\POS[]+<.7pc,0pc>\drop{s}
  }
\end{array}\]}
The production above creates a fresh replica of $f$ at $l'$ and assigns
to it the manager at $g'$; notice that the two instances of $f$ share
the state $s$.

Replication can also trigger a new instance of $f$ that acts on a copy
of the state original state as described in the production of
page~\pageref{page:prod} where $f$ must notify to its state to
duplicate itself and connect the new copy on $s'$.
Hence, the state connected to $s$ duplicate itself on the node $s'$
when the action complementary to $\overline{\gbang}$ is received, as
stated below.

{\small\[\begin{array}{c|c}
  \xymatrix@R1pc{
     *+[F]{\store}\nar|-{\gbang\conf{s'}}[rr] & & \bullet\POS[]+<.7pc,0pc>\drop{s}
  }
  \hspace{1cm}&\hspace{1cm}
  \xymatrix@R1pc{
     *+[F]{\store}\nar[r] & \bullet\POS[]+<.7pc,0pc>\drop{s}
     & \bullet\POS[]+<-.7pc,0pc>\drop{s'}
     & *+[F]{\store}\nar[l]
  }
\end{array}\]}

\paragraph{Component killing}
Components are killed using the following production:

{\small\[\begin{array}{c|c}
  \xymatrix@R1pc{
    & & \bullet\POS[]+<-.7pc,0pc>\drop{l}
    \\
    \bullet\POS[]+<-.7pc,0pc>\drop{g}
    & & *+[F]{f}
    \nar|-{\gkill \conf{}}[ll]
    \nar[u]
    \nar[rr]
    & & \bullet\POS[]+<.7pc,0pc>\drop{s}
  }
  \hspace{1cm}&\hspace{1cm}
  \xymatrix@R1pc@C1pc{
    & & \bullet\POS[]+<-.7pc,0pc>\drop{l}
    \\
    \bullet\POS[]+<-.7pc,0pc>\drop{g}
    & &
    & & \bullet\POS[]+<.7pc,0pc>\drop{s}
  }
\end{array}\]}
stating that $f$ disappears when its corresponding $AM$ sends a $\gkill$
signal.

  \section{Synchronizing productions}
\label{sec:hgex}
The operational semantics of SHR is illustrated through an example
that highlights the following steps:
\smallskip
\begin{compactenum}
\item individuate the adjacent tentacles labeled by compatible conditions;
\item determine the synchronizing productions and replace the (instances of)
  edges on their lhs with the hypergraphs on their rhs;
\item fuse the nodes that are equated by the synchronizations.
\end{compactenum}
\smallskip
Let us apply the previous steps to show how migration works in a
situation represented by the following hypergraph

\[\xymatrix@R1pc@C3pc{
  & &  *+[F]{AM}
  \nar@(l,ur)[dll]|-{\overline{\gstart\store}\conf{g,l_1,s_1}}
  \nar@(r,ul)[rr]
  & \bullet\POS[]+<.5pc,0pc>\drop{l} & \bullet\POS[]+<.5pc,0pc>\drop{l_1}
  \\
  \bullet\POS[]+<-.7pc,0pc>\drop{g}
   & & *+[F]{f}
          \nar|-{\gstart\store\conf{g',l',s'}}[ll]
          \nar@(u,d)[ur]
          \nar[r]
   & \bullet\POS[]+<0pc,.5pc>\drop{s} & *+[F]{\store}\nar[l]
}\]

where component $f$ is running at $l$ and shares $g$ with a manager $AM$
located at $l_1$.
For brevity, tentacles are decorated with the conditions
triggering the rewriting (step 1).
Indeed, the tentacles of $AM$ and of $f$ incident on node $g$ yield compatible
output and input conditions respectively so that $AM$ orders $f$ to migrate
to $l_1$ and to use the store at $s_1$ while staying connected to $g$.

Productions synchronisation consists in replacing the occurrences
of the edges on the lhs with the hypergraphs specified in the rhs of the
productions and applying the node fusions obtained by the node communicated.
For instance, in the previous example the synchronising productions are the
$\gstart\store$ production of $f$ given in Sec.~\ref{shrgrid:sec} and the
production of $AM$ whose lhs and rhs consist of $AM$ connected to $g$ and
$l_1$ (step 2).
Hence, after the synchronization, the node fusions $g' = g$, $l' = l_1$ and
$s' = s_1$ are applied (step 3), so that the hypergraph is rewritten
as

\[\xymatrix@R1pc@C3pc{
  & &  *+[F]{AM}
  \nar@(l,ur)[dll]
  \nar@(r,ul)[rr]
  & \bullet\POS[]+<.5pc,0pc>\drop{l} & \bullet\POS[]+<.5pc,0pc>\drop{l_1}
  \\
  \bullet\POS[]+<-.7pc,0pc>\drop{g}
   & & *+[F]{f}
          \nar[ll]
          \nar@(u,d)[urr]
          \nar@(r,dl)[rrr]
   & \bullet\POS[]+<0pc,.5pc>\drop{s} & *+[F]{\store}\nar[l]
   & \bullet\POS[]+<0pc,.5pc>\drop{s_1}& *+[F]{\store}\nar[l]
}\]

Let us remark that $l$, $\store$ and $s$ remains in the final hypergraph.
In fact they should not be removed because other edges can be allocated
on $l$ or access $\store$. 

The intuitive description of SHR given in this section suggests the following
design style and execution style:
\smallskip
\begin{compactitem}
\item assign an edge to each component and specify their productions;
\item represent the system as a hypergraph;
\item decorate the tentacles with the synchronisation conditions;
\item synchronize the productions until possible.
\end{compactitem}
\smallskip
It is worth remarking that, unlike other semantical frameworks (e.g., process
calculi), in SHR synchronisation conditions may require more than two
(productions of) components to be synchronised.
This actually depends on the synchronisation policy at hand.
For instance, in the migration rewriting described in this section, it is
possible to use broadcast interactions on the node $g$ so that \emph{all}
the components connected on $g$ will move at $l'$ when the productions
are synchronised.

  \section{Conclusions}
\label{rw:sec}
The SHR 
is one of the modelling and theoretical frameworks of the {\sc
Sensoria} project~\cite{Sensoria}. SHR has been exploited
in~\cite{dfmpt03}\ for managing application level \emph{service level
agreement} (SLA) in a distributed environment; in~\cite{fhlmt06}\
several variants of SHR have been described in a uniform context
showing how this formalism can suitably tackle many programming and
modelling facets of arising in service oriented computing.
In this paper we have shown how SHR can be suitably used to formalise
both adaption operations and the evolution of component-based autonomic
applications. 

In an autonomic component model, as GCM, the $AM$ relies on adaptation
operation to realise its management policy (during the
\emph{execution} phase of the autonomic cycle). In general, a
proper formalisation of adaptation operations (possibly involving
the $AM$ itself), may help
\smallskip
\begin{compactitem}
\item to develop correct and effective management policies providing the
  developer with a precise description of the effect (semantics)
  adapation operations; 
\item to formally prove the proprierties of a single adaptations or sequences of
  them;
\item to formally prove local or global invariants characterising the
  evolution of assemblies of autonomic components;
\item to guide developers of the component model itself to establish
  the effectiveness of provided adaptation operations (e.g. the
  completeness of the set of operations with respect to a given
  planned evolution of the assembly).  
\end{compactitem}
\smallskip
This work covers the first of these items, thus represents a first
step in this roadmap, whereas successive items can be enumerated in
the future work. The presented adaptation operations are currently
implemented in the reference implementation of GCM (developed within
the GridCOMP STREP project \cite{gridcomp-web}); their effectiveness
in managing the QoS of grid applications is reported in
\cite{orc:pdp:08}.

\section{Acknowledgments}
The final publication is available at springerlink.com:\\

M.~Aldinucci and E.~Tuosto.
\newblock Towards a formal semantics for autonomic components.
\newblock In T.~Priol and M.~Vanneschi, editors, {\em From Grids To Service and
  Pervasive Computing (Proc. of the CoreGRID Symposium 2008)}, CoreGRID, pages
  31--45, Las Palmas, Spain, Aug. 2008. {Springer}. DOI: 10.1007/978-0-387-09455-7\_3



\end{document}